\documentclass[12pt]{article}
\usepackage{amssymb}
\usepackage{amsfonts, amsmath}
\usepackage{bm}
\usepackage{epsfig}
\usepackage{enumerate}
\usepackage{booktabs}
\usepackage[top=2.5cm,bottom=2.5cm,left=2.5cm,right=2.5cm]{geometry}    
\def\registered{\ooalign{\hfil\raise .00ex\hbox{\scriptsize R}\hfil\crcr\mathhexbox20D}}
\def\tm{\leavevmode\hbox{$\rm {}^{TM}$}}
\newcommand{\vect}{\boldsymbol}
\usepackage{xcolor}

\begin{document}
\title{A Fourier finite volume approach for the optical inverse problem of quantitative photoacoustic tomography}

\author{David J.\ Chappell}

\author{David J.\ Chappell\\
School of Science and Technology,\\
Nottingham Trent University,\\
Clifton Campus,\\
Clifton Lane,\\
Nottingham, UK\\
NG11 8NS\\
david.chappell@ntu.ac.uk} \maketitle

\begin{abstract}
A new approach for solving the optical inverse problem of quantitative photoacoustic tomography is introduced, which interpolates between the well-known diffusion approximation and a radiative transfer equation based model.  The proposed formulation combines a spatial finite volume scheme with a truncated Fourier expansion in the direction variable for the radiative transfer equation. The finite volume scheme provides a natural and simple approach for representing piecewise constant image data modelled using transport equations.  The truncated Fourier expansion in the direction variable facilitates the interpolation between the diffusion approximation at low order, and the full radiative transfer model as the truncation limit $N\rightarrow\infty$.  It is therefore possible to tune the precision of the model to the demands of the imaging application, taking $N=1$ for cases when the diffusion approximation would suffice and increasing the number of terms otherwise.  We will then utilise the non-linear optimisation functionality of Matlab to address the corresponding large-scale nonlinear inverse problem using gradient based quasi-Newton minimisation via the limited memory Broyden-Fletcher-Goldfarb-Shanno algorithm.  Numerical experiments for two test-cases of increasing complexity and resolution will be presented,  and the effect of logarithmically rescaling the problem data on the accuracy of the reconstructed solutions will be investigated.  We will focus on cases where the diffusion approximation is not sufficient to demonstrate that our approach can provide significant accuracy gains with only a modest increase in the number of Fourier terms included.  
\end{abstract}

\section{Introduction}

Photoacoustic tomography (PAT) is an emerging imaging modality with the potential to provide high-resolution images based on optical absorption \cite{LW21}. Nanosecond pulses of laser light are used to obtain the photoacoustic effect in biological tissues whereby an acoustic wave is emitted, which is then detected using an array of ultrasound transducers. As a consequence, there are two parts to the image reconstruction process in quantitative PAT (QPAT): firstly,  an acoustic inverse problem must be solved to reconstruct the acoustic pressure distribution arising from the acoustic waves emitted due to the photoacoustic effect.  Secondly,  an optical inverse problem is solved to reconstruct the chromophore concentrations from the pressure distribution provided by the first inverse problem.  

The first problem of providing a conventional PAT image of the initial pressure distribution is well studied and has become one of the largest research areas in biophotonics during the last twenty years, see for example \cite{LW21, KK08, TZC10} and references therein. This work is instead concerned with the optical inverse problem to determine the chromophore concentration, which aims to combine the large contrast of optical parameters with the high resolution capabilities of ultrasonic waves \cite{BR12}.  QPAT offers promising avenues towards image enhancement through tuning the excitation wavelength to a peak in the absorption spectrum of a particular chromophore, as well as the ability to quantify the concentrations of externally administered contrast agents \cite{STCA13}.  Chromophore concentrations are linearly related to the optical absorption coefficient $\mu_a$ and so in this work we focus on recovering $\mu_a$ through solving an ill-posed, large-scale nonlinear inverse problem.

One of the most common approaches to QPAT (as well as more broadly in optical tomography) is to use the diffusion approximation to the radiative transfer equation (RTE) \cite{BR12, SA99, BR11, TPCKA13}, which relies on the assumption that the propagation of light throughout a tissue is near-isotropic. A major advantage of the diffusion approximation is in terms of model reduction and computational efficiency, since the fluence can be determined directly as the solution of the diffusion equation (or modified Helmholtz equation at steady state) without the need to first compute the radiance, which has both spatial and directional dependence.  However,  in regions close to light sources the propagation of light through biological tissues is highly anisotropic.  Since light sources in QPAT are applied at the tissue surface and these regions are of great interest in QPAT, then using the diffusion approximation can have an adverse effect on the quality of the reconstructed image \cite{STCA13}.  For this reason, some authors have proposed to use a computationally costly but also more accurate model based on the RTE \cite{STCA13, YSJ09}, or alternatively to attempt to get the best of both worlds by constructing hybrid schemes based on the coupling of an RTE based model close to light sources with a diffusion approximation based model elsewhere \cite{TCKA12, TPCA17}. 

Despite being a very popular method for modelling light propagation in tissues \cite{ZL13}, Monte-Carlo models have only relatively recently been employed to model the light propagation within the optical inverse problem of QPAT \cite{LLMPT20,  HPAT22, HPAT23}. The differential equation based models (RTE/diffusion approximation) described above have typically been preferred due to the slow convergence of Monte Carlo and the necessity to recode Monte Carlo techniques for each new absorption distribution, making them particularly unsuitable for the iterative reconstruction algorithms typically used in QPAT.  In addition, the lack of governing differential equations provides a barrier to their implementation in terms of computing the gradients required by the iterative optimisation algorithms for the nonlinear inverse problem.  However, recent progress has seen some of these disadvantages diminished such as the perturbation Monte Carlo method \cite{LLMPT20} which enables forming gradients for the solution of the inverse problem. Furthermore, an adaptive optical Monte Carlo method has recently been proposed in order to achieve sufficient accuracy levels with a reduced computational burden \cite{HPAT22, HPAT23}.

In this work, we develop a numerical approach for the optical inverse problem of QPAT based on the steady-state RTE, that includes the diffusion approximation as a low-order implementation.  \textcolor{black}{We therefore expect our approach to be efficient (only requiring a low-order implementation) for cases where the diffusion approximation is either valid or close to being valid in some sense.} The image reconstruction in QPAT involves finding spatially piecewise constant quantities with jump discontinuities (at boundaries between different biological media) within pixels (or voxels) from piecewise constant data and therefore we base our spatial discretisation of the RTE on a piecewise constant spatial basis expansion.  Since the RTE is a transport equation where the direction of flow is explicitly known as an independent variable in the model, the Finite Volume Method (FVM) provides a relatively straightforward approach for the spatial discretisation and leads to a stable numerical scheme when combined with an upwind approximation for the numerical flux. Finite volume method discretisations are also advantageous for transport models since they preserve conserved quantities (mass, momentum, energy), and for problems with discontinuous coefficients such as the optical parameters reconstructed in QPAT \cite{EGH00}. The FVM has also previously been applied (alongside a discrete ordinate discretisation in the direction variable) in the context of frequency domain optical tomography \cite{RBH06}. Inspired by Fourier expansion based analytical solution techniques for the RTE in both layered \cite{LK12} and infinitely extended scattering media \cite{LK11}, we will base our directional discretisation on a Galerkin projection onto a truncated Fourier expansion. We limit this study to two-dimensional regions, but note that the corresponding directional discretisation for three-dimensional problems would be based on truncated spherical harmonics. Furthermore, these expansions can also be used to derive simplifications to the RTE model known as the $P_N$ approximations \cite{SA99}, for which the $N=1$ case at steady state corresponds to the well-known diffusion approximation (also under some assumptions on the source term that will be satisfied in QPAT where the source arises from the boundary as opposed to within the domain).   Our approach will therefore provide results that are equivalent to those given by the $P_N$ approximations after spatial discretisation via the FVM.

The nonlinear optical inverse problem will be solved using the quasi-Newton minimisation scheme introduced in Ref.~\cite{STCA13} together with the discretisation schemes for the RTE described above. This scheme is advantageous for problems involving large data sets, such as those arising in QPAT, since it only requires gradient information in order to approximate the Hessian matrix using the limited memory Broyden-Fletcher-Goldfarb-Shanno (LBFGS) algorithm. We make use of the Matlab optimization toolbox, which includes these algorithms within the \texttt{fminunc} function \cite{MW24}. Since the dynamic range of the measured light intensities in QPAT can be very large, we consider the use of logarithmically rescaled data in order to improve the accuracy of our reconstructions as proposed in Ref.~\cite{TCKA12}.  

The paper is structured as follows. In the next section we outline the formulation of the problem in terms of a model based on the steady-state RTE and introduce the related quantities of interest. In Section \ref{sec:FVM} we detail the discretisation of the model using the FVM in space and a Fourier based approach in direction.  Section \ref{sec:InvProb} then describes the methodology employed for numerically solving the nonlinear inverse problem generated by the discretisation from section \ref{sec:FVM}, including the details of how the process would differ for logarithmically scaled problem data.  In Section \ref{sec:numerics} we demonstrate the effectiveness of the methods detailed in sections \ref{sec:FVM} and \ref{sec:InvProb} for reconstructing the optical absorption and scattering coefficients in both a small-scale and simple initial test, as well as a larger scale phantom-like image.  Finally, we present our conclusions in Section \ref{sec:conc}.

\section{Problem formulation}\label{sec:ProbDef}

Consider a domain $\Omega\subset\mathbb{R}^2$ that represents a region of biological tissue to be imaged. Light propagation in turbid media can be modelled throughout $\Omega$ in terms of the time-integrated radiance $\phi:\Omega\times S^1\rightarrow\mathbb{R}_+$, where $S^1$ is the unit circle which represents the directional dependence of $\phi$ and $\mathbb{R}_+$ is used to denote non-negative real values.  The time-integrated radiance $\phi$ satisfies the steady-state RTE:
\begin{equation}\label{RTE}
\left(\hat{\mathbf{s}}\cdot\nabla +\mu_a(\bm{x})+ \mu_s(\bm{x})\right)\phi(\bm{x},\hat{\mathbf{s}})-\mu_s(\bm{x})\int_{S^1} \Theta(\hat{\mathbf{s}},\hat{\mathbf{s}}')\phi(\bm{x},\hat{\mathbf{s}}')\mathrm{d}\hat{\mathbf{s}}'=q(\bm{x},\hat{\mathbf{s}}),
\end{equation}
where $\mu_a$ and $\mu_s$ are the absorption and scattering coefficients, respectively, which characterise the propagation of light through $\Omega$ and represent the proportion of light energy that is absorbed or scattered per unit length.  The right hand side term $q:\Omega\times S^1\rightarrow\mathbb{R}_+$ denotes a source of light and $\Theta:S^1\times S^1\rightarrow\mathbb{R}_+$ is the scattering phase function.  We consider the RTE (\ref{RTE}) alongside transparent boundary conditions on $\Omega$, meaning that all light energy reaching the boundary $\Gamma$ simply exits the domain and does not return.  The only light entering $\Omega$ at the boundary arises from boundary source contributions $\phi_0:\Gamma\rightarrow\mathbb{R}_+$. Mathematically, these conditions may be expressed as
\begin{equation}\label{eq:bc}
\phi(\bm{x},\hat{\mathbf{s}})=\phi_0(\bm{x},\hat{\mathbf{s}}), \:\: \bm{x}\in\Gamma,\:\: \hat{\mathbf{s}}\cdot\hat{\mathbf{n}}<0,
\end{equation}
where $\phi_0(\bm{x},\hat{\mathbf{s}})=0$ if $\bm{x}$ is on a source-free region of the boundary and $\hat{\mathbf{n}}$ is the outward unit normal vector at $\bm{x}$.  

The optical inverse problem of QPAT is to recover $\bm{\mu}=[\mu_a, \mu_s]^T$ from the acoustic pressure distribution $p_0$ in $\Omega$, which is obtained from solving the acoustic inverse problem of QPAT.  We will assume that the Gr\"{u}neisen parameter $\gamma$, which connects $p_0$  to the optical energy density $U_{\bm{\mu}}$ via $p_0=\gamma U_{\bm{\mu}}$,  is known throughout $\Omega$. The optical inverse problem is then equivalent to reconstructing $\bm{\mu}$ from the energy density 
\begin{equation}\label{eq:EnDen}
U_{\bm{\mu}}(\bm{x})=\mu_a(\bm{x})\Phi(\bm{x}), 
\end{equation}
where $\Phi$ denotes the total optical energy, or fluence, and is given by
\begin{equation}\label{eq:fluence}
\Phi(\bm{x})=\int_{S^1} \phi(\bm{x},\hat{\mathbf{s}})\mathrm{d}\hat{\mathbf{s}}=\int_{-\pi}^\pi \phi(\bm{x},\hat{\mathbf{s}})\mathrm{d}\theta,
\end{equation}
with $\hat{\mathbf{s}}=[\cos\theta,\sin\theta]^T$. The chromophore concentrations are linearly related to $\mu_a$, and can therefore be obtained from $\mu_a$ provided that all contributing chromophore types are known. Thus the quality of the image reconstruction in QPAT crucially depends on the accuracy of the solution for $\mu_a$.

A common choice for the phase function $\Theta$ with widespread applications including in astrophysics,  atmospheric science and biological media \cite{HG41, HASS03} is the Henyey-Greenstein phase function, which takes the form
\begin{equation}\label{HG2D}
\Theta(\hat{\mathbf{s}},\hat{\mathbf{s}}') = \frac{1-g^2}{2\pi(1+g^2-2g(\hat{\mathbf{s}}\cdot\hat{\mathbf{s}}'))}
\end{equation}
for the case of two (spatial) dimensions.  Here $g\in(-1,1)$ is usually referred to as the anisotropy factor and relates to the mean cosine of the angle 
\[\alpha=\cos^{-1}(\hat{\mathbf{s}}\cdot\hat{\mathbf{s}}')\]
 between the incident and scattered light.  A convenient feature of the Henyey-Greenstein phase function is that it can be expanded in a Fourier series with simple coefficients \cite{HASS03}. In particular, the Fourier expansion of (\ref{HG2D}) takes the form    
\begin{equation}\label{HG2Dexpansion}
\Theta(\hat{\mathbf{s}},\hat{\mathbf{s}}') = \frac{1}{2\pi}\sum_{n=-\infty}^\infty g^{|n|}e^{\mathrm{i}n\alpha}.
\end{equation}
We note that in three dimensions, the Henyey-Greenstein phase function also has an equally simple (Legendre series) expansion in terms of spherical harmonics \cite{HASS03}.  In the next section, we will make use of Fourier expansions alongside a Galerkin projection to discretise the directional dependence of the two dimensional RTE, noting that the extension to three-dimensions would correspondingly utilise Legendre series expansions.
 
\section{Finite volumes and Fourier expansions: Discretisation of the RTE}\label{sec:FVM}
We apply the FVM to (\ref{RTE}) by dividing $\Omega$ into $M$ \textcolor{black}{rectangular} pixels (usually called control volumes in the FVM literature) $\Omega_j$, $j=1,2,\ldots,M$ and assuming that 
\begin{equation}\label{eq:ass1}
\phi(\bm{x},\hat{\mathbf{s}})\approx\frac{1}{\sqrt{2\pi}}\sum_{n=-\infty}^\infty \phi^j_n e^{\mathrm{i}n\theta},
\end{equation}
for $\bm{x}\in\Omega_j$ with $j=1,2,\ldots,M$ \textcolor{black}{and $\hat{\mathbf{s}}=[\cos\theta,\sin\theta]^T$,  as before.} We likewise expand the boundary data $\phi_0$ (see (\ref{eq:bc})) in an identical form to (\ref{eq:ass1}) and label the Fourier coefficients $\phi_n^{j,0}$ when $\bm{x}\in\Gamma\cap\Gamma_j$, with $\Gamma_j$ denoting the boundary of the pixel $\Omega_j$, that is
\[
\phi_0(\bm{x},\hat{\mathbf{s}})\approx\frac{1}{\sqrt{2\pi}}\sum_{n=-\infty}^\infty \phi^{j,0}_n e^{\mathrm{i}n\theta}.
\]
Considering the integral of equation (\ref{RTE}) over $\bm{x}\in\Omega_j$ for some $j=1,2,\ldots,M$ and applying Green's first identity yields
$$\sum_{n=-\infty}^\infty  \frac{e^{\mathrm{i}n\theta}}{\sqrt{2\pi}}\left(\int_{\Gamma_j}\left(\hat{\mathbf{s}}\cdot\hat{\mathbf{n}}\right)(\mathcal{P}{\phi}^j_n)(\bm{x},\hat{\mathbf{s}})\mathrm{d}\Gamma_j+|\Omega_j|\left(\mu^j_a+ \left(1- g^{|n|}\right)\mu^j_s\right) \phi^j_n \right)=0,$$
where we have taken $q\equiv0$ since the source terms in QPAT only enter at the boundaries via the boundary integral term on the left hand side.  We have also introduced the notation $|\Omega_j|$ to denote the area of the pixel $\Omega_j$ and $\mu_a^j$, $\mu_s^j$ to denote the piecewise constant values of  $\mu_a(\bm{x})$,  $\mu_s(\bm{x})$, respectively,  when $\bm{x}\in\Omega_j$. 
The integral term in the RTE (\ref{RTE})  has been evaluated by introducing the expansion (\ref{HG2Dexpansion}) and applying orthogonality. The notation $\mathcal{P}\phi_j^n$ relates to the projection of the radiance $\phi$ onto the pixel boundary $\Gamma_j$, where the approximation (\ref{eq:ass1}) means that $\phi(\bm{x},\hat{\mathbf{s}})$ is not well defined since, in general, there will be a jump discontinuity when $\bm{x}\in\Gamma_j$. A simple choice for $\mathcal{P}$ that leads to a stable numerical scheme is the upwind scheme defined by
\[(\mathcal{P}\phi_n^j)(\bm{x},\hat{\mathbf{s}})=\left\{\begin{array}{lll}\phi_n^j & \mathrm{if} & \hat{\mathbf{s}}\cdot\hat{\mathbf{n}}>0,\\
\phi_n^{j+} & \mathrm{otherwise,} & \end{array}\right.\]
where $\phi_n^{j+}$ denotes the corresponding coefficient from (\ref{eq:ass1}) associated to the pixel adjacent to $\Omega_j$ at $\bm{x}\in\Gamma_j$ (assuming $\bm{x}$ is not a vertex). In the case where $\Gamma_j\subset\Gamma$ and hence there is no adjacent pixel, then we apply the boundary condition (\ref{eq:bc}) to give $\phi_n^{j+}=\phi_n^{j,0}$.

\begin{figure}[h!]
\centering
\includegraphics[trim={3.8cm 4cm 4cm 4cm},clip,width=0.45\textwidth]{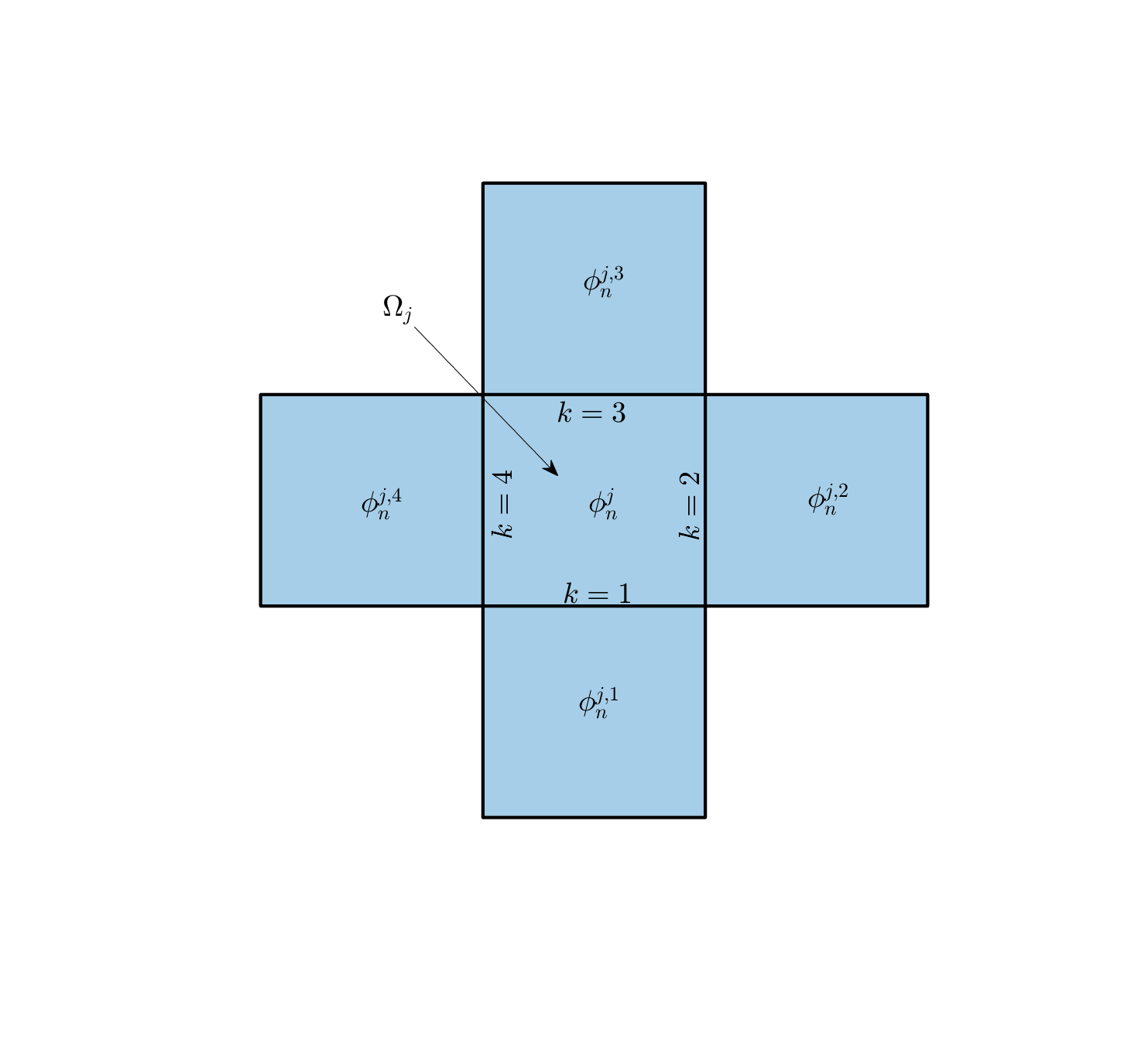}
\caption{\label{fig:Grid} Pixel $\Omega_j$ and adjacent pixels connected by edges $\Gamma_k$, $k=1,2,3,4$. The labelling shows the expansion coefficient from (\ref{eq:ass1}) associated to each pixel.}
\end{figure}

We now apply a finite dimensional Galerkin projection in $\theta$ by truncating the Fourier expansion at $2N+1$ terms, multiplying through by $e^{-\mathrm{i}m\theta}/\sqrt{2\pi}$, for some $m=-N,\ldots,N$ and integrating over $S^1$ as follows
$$
\left(\sum_{n=-N}^N   \int_{-\pi}^\pi \frac{e^{\mathrm{i}(n-m)\theta}}{2\pi}\int_{\Gamma_j}\left(\hat{\mathbf{s}}\cdot\hat{\mathbf{n}}\right)(\mathcal{P}{\phi}^j_n)(\bm{x},\hat{\mathbf{s}})\mathrm{d}\Gamma_j\mathrm{d}\theta\right)+|\Omega_j|\left(\mu^j_a+ \left(1- g^{|m|}\right)\mu^j_s\right) \phi^j_m =0.
$$
\textcolor{black}{Note that orthogonality has been applied to evaluate the $\theta$ integral exactly in the second term.} The first term may be evaluated by splitting the boundary integral into a sum of integrals along the (four) edges of the pixel boundary $\Gamma_j$ and evaluating the scalar product, the edge projection and the spatial integral accordingly as follows
\begin{align}
\begin{split}\label{eq:FVM}
&\sum_{n=-N}^N \int_{-\pi}^\pi \frac{e^{\mathrm{i}(n-m)\theta}}{2\pi}\left(w_j|\sin\theta|\left(\phi_n^j-\phi_n^{j,1}H(\theta)-\phi_n^{j,3}H(-\theta)\right)\ldots\frac{}{}\right.\\
&\left.+h_j|\cos\theta|\left(\phi_n^j-\phi_n^{j,2}H\left(|\theta|-\frac{\pi}{2}\right)-\phi_n^{j,4}H\left(\frac{\pi}{2}-|\theta|\right)\right)\right)\mathrm{d}\theta
 +|\Omega_j|\left(\mu^j_a+ \left(1- g^{|m|}\right)\mu^j_s\right) \phi^j_m =0.
\end{split}
\end{align}
Here we have introduced the notation $w_j$ and $h_j$ for the width and height of the pixel $\Omega_j$, respectively, and thus $|\Omega_j|=w_jh_j$. We also make use of the Heaviside step function $H$ to restrict the integral to the appropriate range as specified by the upwind scheme.  The notation $\phi_n^{j,k} $ refers to the coefficient from (\ref{eq:ass1}) associated to the pixel joined to $\Omega_j$ along the $k^\mathrm{th}$ edge of $\Gamma_j$, where the edge numbering is oriented anticlockwise and starts from the bottom edge  as shown in Fig \ref{fig:Grid}.  \textcolor{black}{We anticipate that a relatively small value for $N$ will be sufficient in many examples,  with $N=1$ corresponding to the well-known diffusion approximation of the RTE and larger choices of $N$ incorporating more of the radiative transfer dynamics.}

The remaining integral appearing in (\ref{eq:FVM}) is reasonably simple to evaluate, treating  $|n-m|=1$ as a special case since the general result becomes singular here.  As a consequence, the assembly of the scheme (\ref{eq:FVM}) into a matrix vector system is relatively straightforward requiring no numerical integration. The non-zero components of the right hand side vector are formed by applying the boundary condition to set $\phi_n^{j,k}=\phi_n^{j,0}$ whenever $\Gamma_j\subset\Gamma$ and moving those known terms over to the right hand side. The matrix vector system for each pixel $\Omega_j$ is therefore formed by writing out the system (\ref{eq:FVM}) for each $m=-N,\ldots,N$ with the sum being performed via the matrix vector multiplication, and then the global system is assembled by writing out the system for each pixel in turn. For compactness in the sequel we write this matrix vector equation as 
\begin{equation}\label{eq:LinSys}
A\bm{\phi} = \bm{b},
\end{equation} 
where \textcolor{black}{$\bm{\phi}=[\phi^1_{-N}, \phi^1_{-N+1}, \ldots, \phi^1_{N},\phi^2_{-N},\ldots,\ldots,\phi^{M}_{-N},\ldots,\phi^M_{N}]^T$} is a vector of length $(2N+1)M$ and $A$, $\bm{b}$ are assembled from (\ref{eq:FVM}) as described above.  Once $\bm{\phi}$ has been computed, the fluence may be computed via (\ref{eq:fluence}) and thus the optical energy density is obtained from (\ref{eq:EnDen}). Substituting the approximation (\ref{eq:ass1}) into (\ref{eq:fluence}) simplifies the evaluation of the fluence $\Phi(\bm{x})=\Phi_{j}$ for $\bm{x}\in\Omega_j$, $j=1,\ldots,M$ as follows
\begin{equation}\label{eq:forprobsolj}
\Phi_{j}=\frac{1}{\sqrt{2\pi}}\sum_{n=-\infty}^\infty \phi^j_n\int_{-\pi}^\pi  e^{\mathrm{i}n\theta}\mathrm{d}\theta=\sqrt{2\pi}\phi^j_0 .
\end{equation}
Note that it is often convenient to express (\ref{eq:forprobsolj}) for all $j=1,\ldots,M$ in a single equation of the form 
\begin{equation}\label{eq:forprobsolall}
\bm{\Phi}=T\bm{\phi},
\end{equation}
where $\bm{\Phi}=[\Phi_{j}]_{j=1,\ldots,M}$ and $T$ is a sparse $M\times(2N+1)M$ matrix whose non-zero entries \textcolor{black}{are all} $\sqrt{2\pi}$\textcolor{black}{, and} are found only in the positions $(j,(2j-1)N+j)$ for $j=1,\ldots,M$.  In the next section we describe an iterative scheme for solving the optical inverse problem which involves the repeated evaluation of the optical energy density (\ref{eq:EnDen}) using (\ref{eq:forprobsolall}) with updated values of $\mu_a^j$ and $\mu_s^j$ for $j=1,\ldots,M$ in order to minimise a prescribed objective function.

\section{Adjoint assisted quasi-Newton minimisation for the optical inverse problem}\label{sec:InvProb}
We now outline an efficient quasi-Newton minimisation scheme based on the LBFGS algorithm and implemented via the \texttt{fminunc} function from the Matlab optimisation toolbox \cite{MW24}. We supply gradient information during each iteration in order to approximate the Hessian matrix to update the values of $\mu_a^j$ and $\mu_s^j$ for $j=1,\ldots,M$. This gradient information will be obtained from an adjoint problem for the RTE (\ref{RTE}) as proposed in Ref.~\cite{STCA13}.  The recovery of both the absorption and scattering coefficients using a single PAT image is non-unique \cite{BU10}. Here this issue is circumvented through the use of multiple source projections taken from different parts of the boundary of $\Omega$.

We seek approximate solutions $\hat{\bm{\mu}}\approx\bm{\mu}$ to the optical inverse problem through minimising the least-squares error functional with data $U^p_{\mathrm{dat}}$ obtained from multiple projections $p=1,2,\ldots,P$ as follows
\begin{equation}\label{eq:ErFunc}
\hat{\bm{\mu}}=\arg\min_{\bm{\mu}}\left(\frac{1}{2}\sum_{p=1}^P\int_{\Omega}(f(U^p_{\mathrm{dat}}(\bm{x})) - f(U^p_{\bm{\mu}}(\bm{x})))^2\mathrm{d}\bm{x}+\mathcal{R}(\bm{\mu})\right).
\end{equation}
The function $f$ is used to define the scaling of the optical energy density,  which may be useful for lossy turbid media where the dynamic range of the data can span many orders of magnitude.  We will consider the cases when $f$ is either the natural logarithm function $f=\ln$ or the identity function $f=\mathrm{id}$.  We have also denoted  the optical energy density ({\ref{eq:EnDen}) computed for the $p^{\mathrm{th}}$ source projection as $U^p_{\bm{\mu}}$. The final term on the right side $\mathcal{R}$ is a regularisation penalty term, which is usually necessary to dampen the effect of noise in the data on the reconstructed image.  In this work, we take \textcolor{black}{
\begin{equation}\label{TikReg}
\mathcal{R}(\bm{\mu})=\frac{\alpha}{2}|\nabla\mu_a|^2+\frac{\beta}{2}|\nabla\mu_s|^2,
\end{equation}
corresponding to first-order Tikhonov regularisation with regularisation parameters $\alpha$ and $\beta$ as proposed in Ref. \cite{STCA13}. The gradients are evaluated over the pixel-based grid using standard second-order accurate central difference approximations, taking the pixel centroids as the nodes.} We generate the data $U_{\mathrm{dat}}^p$ synthetically by solving the corresponding forward light transport problem using the ValoMC Monte Carlo toolbox for Matlab \cite{LPT19} using $P$ different source illumination terms. 

Introducing the notation $\mathcal{E}_f$ for the error functional in (\ref{eq:ErFunc}) and writing 
\begin{align*}
\mathcal{E}_f(\vect{\mu})&=\sum_{p=1}^P \frac{1}{2}\int_{\Omega}(f(U^p_{\mathrm{dat}}(\vect{x})) - f(U^p_{\vect{\mu}}(\vect{x})))^2\mathrm{d}\vect{x}+\mathcal{R}(\vect{\mu})\\
&=\sum_{p=1}^P \mathcal{E}_f^p(\vect{\mu}) + \mathcal{R}(\vect{\mu}),
\end{align*}
then the required functional gradients take the form
\[\frac{\partial\mathcal{E}_f}{\partial \mu}=\sum_{p=1}^P \frac{\partial\mathcal{E}_f^p}{\partial\mu} + \frac{\partial\mathcal{R}}{\partial\mu},\]
where $\mu=\mu_a$ or $\mu_s$.  \textcolor{black}{The derivative of $\mathcal{R}$ is given by 
$$ \frac{\partial\mathcal{R}}{\partial\mu}=\left(\alpha\delta_{\mu, \mu_a} + \beta\delta_{\mu,\mu_s}\right)\Delta\mu,$$
where $\delta_{\mu,\mu_{a}}$ and $\delta_{\mu,\mu_{s}}$ are Kronecker deltas and we have assumed that $\mu_s\neq\mu_a$. The evaluation of $\Delta \mu$ is carried out via standard second-order accurate central difference approximations,  as for the gradients previously.} Noting the piecewise constant nature of the data $U^p_{\mathrm{dat}}$ and the FVM solution for the fluence $\Phi$, then the integral required to compute $\mathcal{E}_f^p$ reduces to a sum as follows
\begin{align*}
\mathcal{E}_f^p(\vect{\mu}) &=\frac{1}{2}\int_{\Omega}(f(U^p_{\mathrm{dat}}(\vect{x})) - f(U^p_{\vect{\mu}}(\vect{x})))^2\mathrm{d}\vect{x}\\
&=\frac{1}{2}\sum_{j=1}^M |\Omega_j|(f(U^p_{\mathrm{dat},j}) - f(U^p_{\vect{\mu},j}))^2\\
&=\frac{1}{2}(f(\vect{U}^p_{\mathrm{dat}}) - f(\vect{U}^p_{\vect{\mu}}))^T((f(\vect{U}^p_{\mathrm{dat}}) - f(\vect{U}^p_{\vect{\mu}}))\odot\vect{\Omega}),
\end{align*}
where $U^{p,  j}_{\mathrm{dat}}=U^p_{\mathrm{dat}}(\bm{x})$ for any $\bm{x}\in\Omega_j$,  $j=1,2,\ldots,M$ and $\bm{U}^p_{\mathrm{dat}}=[U^{p, j}_{\mathrm{dat}}]_{j=1,\ldots,M}$ with corresponding notation for $U^p_{\bm{\mu}}(\bm{x})$. We have also introduced $\bm{\Omega}=[|\Omega_j|]_{j=1,\ldots,M}$ and use $\odot$ for the Hadamard product.  Following Ref.~\cite{STCA13}, we may then calculate the gradient of $\mathcal{E}^P$ as follows
\begin{align}\notag
\frac{\partial\mathcal{E}_f^p}{\partial\mu} &=-\left(\frac{\partial f(\vect{U}^p_{\vect{\mu}})}{\partial\mu}\right)^T((f(\vect{U}^p_{\mathrm{dat}}) - f(\vect{U}^p_{\vect{\mu}}))\odot\vect{\Omega})\\\label{funcgradl2}
&=-\left(\frac{\partial f(\vect{\mu}_a\odot\vect{\Phi}^p)}{\partial\mu}\right)^T((f(\vect{U}^p_{\mathrm{dat}}) - f(\vect{U}^p_{\vect{\mu}}))\odot\vect{\Omega}).
\end{align}
Here, we have set $\bm{\mu}_a=[\mu_a^j]_{j=1,\ldots,M}$,  and $\bm{\Phi}^p=[\Phi_j^p]_{j=1,\ldots,M}$ denotes the vector given by the fluence values associated with each of the $M$ pixels for the $p$th source projection.  At this point we proceed differently for the two choices of the function $f$. If $f$ is taken as the identity function in (\ref{funcgradl2}) then we simply obtain
\begin{equation}\label{funcgradunsc}
\frac{\partial\mathcal{E}_{\mathrm{id}}^p}{\partial\mu} =-\left({\bm{\mu}_a}\odot\frac{\partial\bm{\Phi}^p}{\partial\mu}+\delta_{\mu,\mu_{a}}\bm{\Phi}^p\right)^T(\bm{U}^p_{\mathrm{dat}} - \bm{U}^p_{\bm{\mu}})\odot\bm{\Omega}),
\end{equation}
and for the case when $f$ prescribes a logarithmic scaling we apply the laws of logarithms to instead obtain
\begin{equation}\label{funcgradlnsc}
\frac{\partial\mathcal{E}_{\ln}^p}{\partial\mu} =-\left((\bm{\Phi}^p)^{\circ(-1)}\odot\frac{\partial\bm{\Phi}^p}{\partial\mu}+\delta_{\mu,\mu_{a}}{\bm{\mu}_a}^{\circ(-1)}\right)^T((\ln(\bm{U}^p_{\mathrm{dat}}) - \ln(\bm{U}^p_{\bm{\mu}}))\odot\bm{\Omega}).
\end{equation}
Here, the superscript $\circ(-1)$ is used for the Hadamard inverse of vectors with only non-zero (in our case,  strictly positive) entries.

At this point it is convenient to recall the connection between the fluence and the radiance (\ref{eq:forprobsolall}) and also noting that the matrix $T$ is independent of both $\mu_a$ and $\mu_s$ allows us to rewrite
\begin{equation}\label{eq:dPhi}
\frac{\partial\bm{\Phi}^p}{\partial\mu} = T\frac{\partial\bm{\phi}^p}{\partial\mu},
\end{equation}
where $\bm{\phi}^p$ is the vector of radiance values for the $p$th source projection. The right side may then be evaluated by first differentiating (\ref{eq:LinSys}) to give
\[\frac{\partial (A\bm{\phi}^p)}{\partial \mu} =\frac{\partial A}{\partial \mu}\bm{\phi}^p+A\frac{\partial \bm{\phi}^p}{\partial \mu}=0,\]
where the final zero term is a consequence of the right hand side vector in (\ref{eq:LinSys}) being independent of both $\mu_a$ and $\mu_s$. Rearranging and substituting into (\ref{eq:dPhi}) yields
\[\frac{\partial\bm{\Phi}^p}{\partial\mu} = -T A^{-1}\frac{\partial A}{\partial \mu}\bm{\phi}^p,\]
which we the apply in the functional gradient expressions (\ref{funcgradunsc}) and ({\ref{funcgradlnsc}) to obtain
\begin{equation}\label{eq:dE1id}
\frac{\partial\mathcal{E}_{\mathrm{id}}^p}{\partial\mu} 
=\left({\bm{\mu}_a}\odot\left(T A^{-1}\frac{\partial A}{\partial \mu}\bm{\phi}^p\right)-\delta_{\mu,\mu_{a}}\bm{\Phi}^p\right)^T(\bm{U}^p_{\mathrm{dat}} - \bm{U}^p_{\bm{\mu}})\odot\bm{\Omega}),
\end{equation}
and
\begin{equation}\label{eq:dE1ln}
\frac{\partial\mathcal{E}_{\ln}^p}{\partial\mu} 
=\left((\vect{\Phi}^p)^{\circ(-1)}\odot \left(T A^{-1}\frac{\partial A}{\partial \mu}\vect{\phi}^p\right)-\delta_{\mu,\mu_{a}}{\vect{\mu}_a}^{\circ(-1)}\right)^T(\ln(\vect{U}^p_{\mathrm{dat}}) - \ln(\vect{U}^p_{\vect{\mu}}))\odot\vect{\Omega}),
\end{equation}
respectively. Splitting (\ref{eq:dE1id}) into two terms and applying the transpose within the parentheses provides
$$
\frac{\partial\mathcal{E}_{\mathrm{id}}^p}{\partial\mu} =
{\bm{\mu}_a}^T\odot\left((\bm{\phi}^p)^T\frac{\partial A}{\partial \mu}^T(A^{-1})^T T^T\right) ((\bm{U}^p_{\mathrm{dat}} - \bm{U}^p_{\bm{\mu}})\odot\bm{\Omega})-\left(\delta_{\mu,\mu_{a}}\bm{\Phi}^p\right)^T((\bm{U}^p_{\mathrm{dat}} - \bm{U}^p_{\bm{\mu}})\odot\bm{\Omega}),
$$
which may be simplified to 
\begin{equation}\label{eq:FuncGrad}
\frac{\partial\mathcal{E}_{\mathrm{id}}^p}{\partial\mu} 
=(\bm{\phi}^p)^T\frac{\partial A}{\partial \mu}^T\bm{\phi}_{\mathrm{id}}^{p*} -\left(\delta_{\mu,\mu_{a}}\bm{\Phi}^p\right)^T((\bm{U}^p_{\mathrm{dat}} - \bm{U}^p_{\bm{\mu}})\odot\bm{\Omega}),
\end{equation}
upon setting $\bm{\phi}_{\mathrm{id}}^{p*}$ as the solution to the adjoint problem
\begin{equation}\label{eq:AdjProb}
A^T\bm{\phi}_{\mathrm{id}}^{p*}= T^T (\bm{\mu}_a\odot(\bm{U}^p_{\mathrm{dat}} - \bm{U}^p_{\bm{\mu}})\odot\bm{\Omega}).
\end{equation}
A similar procedure may be applied to the logarithmically scaled case, whereby one instead obtains
\begin{equation}\label{eq:FuncGradLn}
\frac{\partial\mathcal{E}_{\ln}^p}{\partial\mu} 
=(\bm{\phi}^p)^T\frac{\partial A}{\partial \mu}^T\bm{\phi}_{\ln}^{p*} -\left(\delta_{\mu,\mu_{a}}{\bm{\mu}_a}^{\circ(-1)}\right)^T((\ln(\bm{U}^p_{\mathrm{dat}}) - \ln(\bm{U}^p_{\bm{\mu}}))\odot\bm{\Omega}),
\end{equation}
where $\bm{\phi}_{\ln}^{p*}$ is the solution to the adjoint problem
\begin{equation}\label{eq:AdjProbLn}
A^T\bm{\phi}_{\ln}^{p*}= T^T ((\bm{\Phi}^p)^{\circ(-1)}\odot(\ln(\bm{U}^p_{\mathrm{dat}}) - \ln(\bm{U}^p_{\bm{\mu}}))\odot\bm{\Omega}).
\end{equation}
As a consequence,  the functional gradients (\ref{eq:FuncGrad}) or (\ref{eq:FuncGradLn}) (with $\mu$ taken to be either $\mu_a$ or $\mu_s$) may be calculated from solving both the forward problem (\ref{eq:LinSys}) and the corresponding  adjoint problem (either (\ref{eq:AdjProb}) or (\ref{eq:AdjProbLn})) for each source projection $p=1,\ldots,P$. The partial derivatives of $A$ are simple to compute from (\ref{eq:FVM}) and take the form of diagonal matrices with $2N+1$ entries along the diagonal per pixel.  For $\partial A/\partial\mu_a$, the $2N+1$ entries for the $j$th pixel are all simply $|\Omega_j|$.  For $\partial A/\partial\mu_s$, the $2N+1$ entries for the $j$th pixel are given as $(1-g^{|n|})|\Omega_j|$ for $n=-N,\ldots, N$.  In the next section, we apply both the unscaled and logarithmically scaled data and their corresponding functional gradients within the Matlab unconstrained optimisation function \texttt{fminunc} in order to minimise the error functional (\ref{eq:ErFunc}),  and thereby reconstruct the absorption and scattering coefficients for two different test images of varying resolution and complexity.

\section{Numerical results and discussion}\label{sec:numerics}
In this section we apply the methodology outlined in sections \ref{sec:ProbDef} to \ref{sec:InvProb} to investigate the optical inverse problem of QPAT for two example images.  In each case the image is located within a square region with $P=4$ source projections given by spatially constant line sources directed perpendicularly into the domain from each boundary edge.  That is in (\ref{eq:bc}) we have 
$$\phi_0(\bm{x},\hat{\mathbf{s}})=\delta(\theta-\theta^p_0),$$
where $\theta^1_0=\pi/2$ corresponds to the source on the lower edge and $\theta_0^p=(p-4)\pi/2$, $p=2,3,4$ corresponds to the subsequent edges listed in anti-clockwise order.  We will first consider a smaller domain (4mm$\times$4mm) with a simple test image to provide a proof of concept for our approach.  The second example features a larger (8mm$\times$8mm) domain with a more complex image based on a modified Shepp-Logan phantom \cite{SL74}. The optical properties in each case were chosen to be within the range typical of biological tissues, including a low scattering outer region and a more highly scattering inner region.  \textcolor{black}{In these examples,  the source projection boundary conditions are chosen to align exactly with the edges of the pixel based finite volume mesh used here.  If the geometry of the domain $\Omega$ was chosen more generally,  then it would be necessary to either use a suitably fine mesh whose edges give a reasonable approximation of the true geometry,  or to generalise the finite volumes to include other geometries that can better represent the boundary $\Gamma$. In the latter case, this would require a generalisation of (\ref{eq:FVM}) to compute the boundary integral(s) for more general geometries.}

Our goal is to reconstruct the absorption and scattering coefficients to see how well we can detect abrupt changes in these values from the corresponding absorbed optical energy density data $U_{\mathrm{dat}}^p$, for $p=1,..,4$. This data was generated using ValoMC \cite{LPT19} to obtain the fluence (\ref{eq:fluence}) and thus the energy density from (\ref{eq:EnDen}). \textcolor{black}{Note that  the ValoMC Monte Carlo toolbox uses the photon packet method \cite{PKJW89},  which is a statistical approach not based on differential equations, and the results are generated on a triangular mesh where the triangles are chosen to correspond to our FVM pixels divided into two along one of the diagonals.  The pixel data are then evaluated by taking the mean of the ValoMC solution values from the two triangles that form each pixel.}  We set the Gr\"{u}neisen parameter $\gamma(\bm{x})=1$ for all $\bm{x}\in\Gamma$ so that for our examples, the optical energy density takes the same values as the acoustic pressure distribution arising in the solution of the acoustic inverse problem of QPAT.  In both examples,  Gaussian noise at a level of $5\%$ of the acoustic pressure was added to the ValoMC results used to provide the synthetic problem data, in order to replicate the measurement noise encountered in practice.  

\subsection{Initial study}
 \begin{figure}[h!]
\centering
\includegraphics[trim={1.5cm 0.5cm 1.5cm 0.5cm},clip,width=0.75\textwidth]{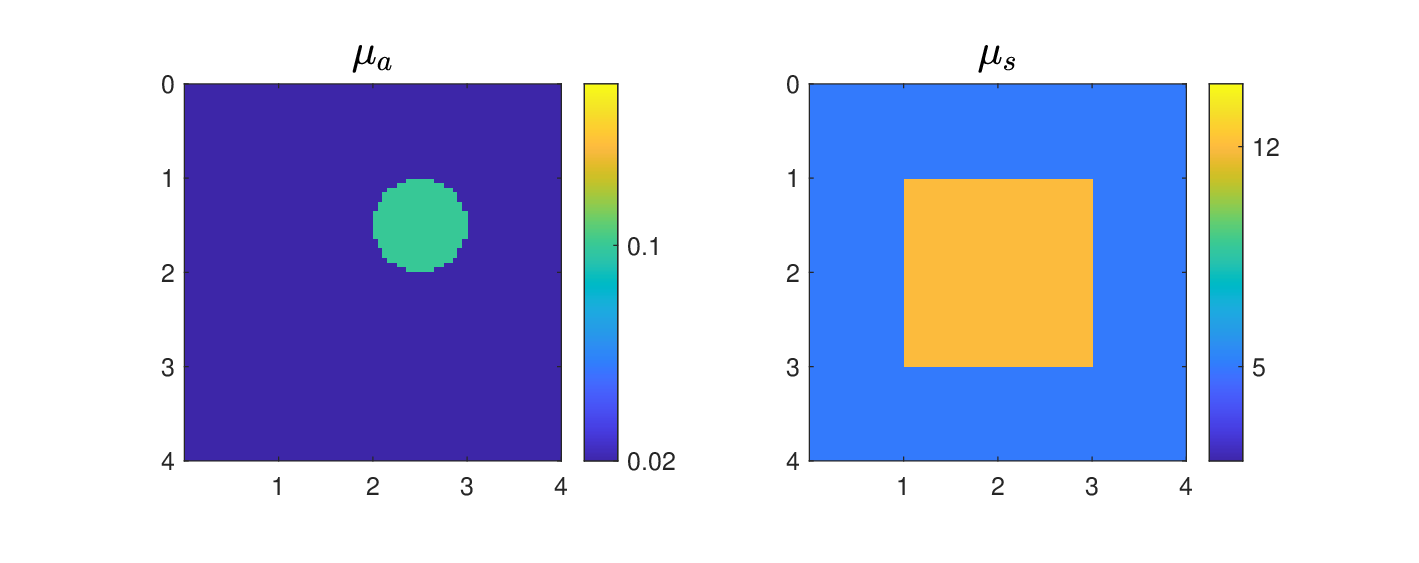}
\caption{\label{fig:GroundTruthEx1} Ground truth values of the absorption (left) and scattering (right) coefficients for the initial study.  The colour axis has been extended for better comparison with the numerically reconstructed absorption and scattering coefficients.}
\end{figure}
We seek to reconstruct the absorption and scattering coefficients in a 4mm$\times$4mm domain where the background values are taken as $\mu_a=0.02\mathrm{mm}^{-1}$, $\mu_s=5\mathrm{mm}^{-1}$ with anisotropy factor $g=0.8$.  \textcolor{black}{These background values are also used to provide constant initial guesses for $\mu_a$ and $\mu_s$ for the iterative quasi-Newton minimisation scheme.} The ground truth values for $\mu_a$ and $\mu_s$ used in the generation of the synthetic data $U^p_{\mathrm{dat}}$ using ValoMC \cite{LPT19} are the piecewise constant values shown in Figure \ref{fig:GroundTruthEx1}. The images here are shown on a grid of 80$\times$80 pixels, which is also the number of the pixels $M=6400$ used for the finite volume discretisation of the inverse problem. We will investigate the influence of changing the number of Fourier terms $N$ on the accuracy of the reconstructions, recalling that the choice $N=1$ corresponds to the commonly used diffusion approximation for the configuration here.  We also investigate whether improved results can be obtained by logarithmically scaling the optical energy density as described in Section \ref{sec:InvProb}.  \textcolor{black}{Regularisation has not been applied for the results of this initial study (that is, we take $\alpha=\beta=0$ for the regularisation parameters in (\ref{TikReg})), since we found that while small improvements could be obtained in the results for $\mu_a$, these were offset by significantly worse results for $\mu_s$. The regularisation parameters also needed to be chosen carefully for each different $N$ value since a fixed choice for all $N$ values would lead to improvements for some results and lower accuracy for others. We will discuss this further in the context of the results below, but we hypothesise that regularisation is not imperative here due to relatively low noise levels and the relatively simplistic and small-scale nature of this initial test. }
 \begin{figure}[h!]
\centering
\includegraphics[trim={4cm 0.5cm 3.5cm 0cm},clip,width=\textwidth]{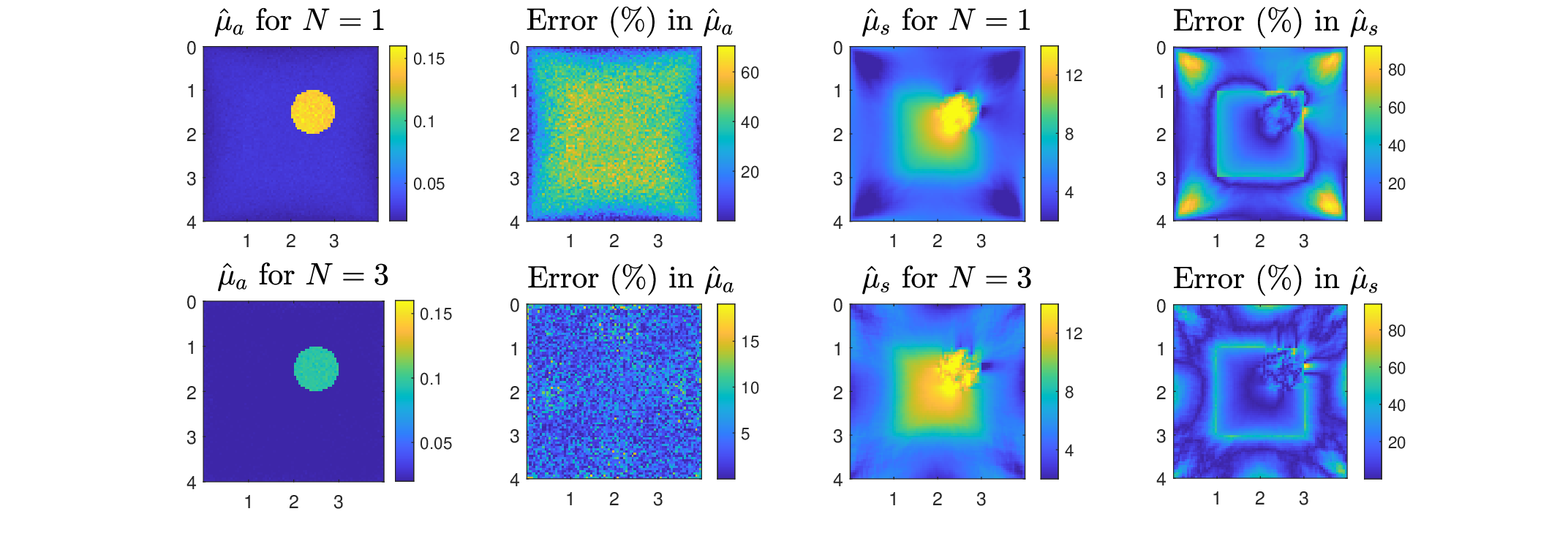}
\caption{\label{fig:UnscaledEx1} Values of the absorption (\textcolor{black}{column 1}) and scattering (\textcolor{black}{column 3}) coefficients reconstructed \textcolor{black}{from 400 iterations of the quasi-Newton algorithm} using unscaled \textcolor{black}{noisy} optical energy density data for the initial study,  with different truncation values $N$ of the Fourier expansion giving the directional approximation. \textcolor{black}{Columns 2 and 4 show the (percentage) reconstruction errors in each pixel corresponding to the adjacent result in columns 1 and 3, respectively.}}
\end{figure}

\begin{table}
\caption{\label{tab:Ex1} Percentage errors (\ref{eq:error}) in the reconstructions of the absorption and scattering coefficients shown in Figure \ref{fig:GroundTruthEx1} \textcolor{black}{after 400 iterations of the quasi-Newton algorithm} for different truncation values $N$ of the Fourier expansion giving the directional approximation, and for both unscaled and logarithmically scaled data. \textcolor{black}{Results are also provided for both data with added Gaussian noise at a level of $5\%$ of the acoustic pressure, as well as data with no added noise.}}
\begin{center}
\begin{tabular}{@{}ccccccccccc}
 & \multicolumn{5}{c}{Unscaled data} & \multicolumn{5}{c}{ Logarithmically scaled data}\\
 & \multicolumn{2}{c}{Noisy} & \multicolumn{2}{c}{Clean} & & \multicolumn{2}{c}{Noisy} &\multicolumn{2}{c}{Clean} & \\  
 \cmidrule(r){2-6}  \cmidrule(r){7-11}
$N$ & $E(\mu_a)$ & $E(\mu_s)$ & $E(\mu_a)$ & $E(\mu_s)$ & Time (s) & $E(\mu_a)$ & $E(\mu_s)$ & $E(\mu_a)$ & $E(\mu_s)$ & Time (s)\\ \hline
1 & 42.6 & 30.3 & 42.6 & 30.1 & 308  & 44.1 & 25.8 & 43.4 & 26.7   & 331 \\
2 & 5.35 & 25.5 & 4.18 & 23.1 & 939  & 5.62 & 21.5 & 4.29 & 21.5   & 1133 \\
3 & 4.93 & 20.2 & 3.52 & 19.0 & 1890 & 3.71 & 16.9 & 1.44 & 17.1   & 2265\\
4 & 4.51 & 20.3 & 2.27 & 17.1 & 3115 & 3.22 & 18.0 & 0.0987 & 16.6 & 3504\\
5 & 4.48 & 21.5 & 1.62 & 18.0 & 5083 & 3.38 & 19.4 & 2.14   & 18.6 & 5319\\
6 & 4.71 & 22.9 & 3.16 & 20.6 & 6625 & 4.07 & 18.7 & 2.70   & 18.5 & 7188
\end{tabular}
\end{center}
\end{table}
Figure \ref{fig:UnscaledEx1} shows a comparison between the absorption and scattering coefficients reconstructed with $N=1$ and $N=3$ Fourier terms \textcolor{black}{after 400 iterations of the quasi-Newton algorithm using unscaled noisy data}.  One observes in both cases that the \textcolor{black}{error in the reconstruction of $\mu_s$ is structured with peaks close to the regions where $\mu_a$ and $\mu_s$ undergo abrupt changes, as well as near the corners of the domain.  In contrast, the error in the reconstruction of $\mu_a$ is less structured,  but increases towards the centre of the domain for $N=1$.}  In comparison to the ground truth shown in Figure \ref{fig:GroundTruthEx1}, one observes a significant improvement in the accuracy for the $N=3$ reconstruction compared to the result with $N=1$,  and this is particularly true for the $\mu_a$ result, which is visually almost identical to the ground truth. These observations are quantified in Table \ref{tab:Ex1}, which lists the percentage errors $E(\mu_a)$ and $E(\mu_s)$ for the reconstructed absorption and scattering coefficients, respectively. These errors are computed via
\begin{equation}\label{eq:error}
E(\mu)=\sqrt{\frac{\sum_{j=1}^{M} (\mu^j - \hat{\mu}^j)^2}{\sum_{j=1}^{M} (\mu^j)^2}}\times100\%
\end{equation}
for $\mu=\mu_a$ or $\mu=\mu_s$.  Note that we have used a superscript $j$ to denote the value of the quantity in pixel $\Omega_j$ and the hat notation to distinguish the reconstructed optical parameter from the ground truth value.  With reference to the results plotted in Figure \ref{fig:UnscaledEx1}, we note that increasing $N$ from 1 to 3 gives a decrease in $E(\mu_a)$ from $42.6\%$ to $4.9\%$ and a decrease in $E(\mu_s)$ from $30.3\%$ to $20.2\%$. We also note that increasing $N>3$ does not provide a significant or consistent decrease in either error value.  

\textcolor{black}{The saturation of the error (rather than a continued monotonic decrease) is expected due to the following factors; firstly, the spatial finite volume grid is fixed and so all results will include an error contribution from the spatial discretisation that will be independent of the choice of $N$.  Secondly,  the noise in the data will contribute to the overall error and we note that in addition to the 5\% additive Gaussian noise, the data will also contain `noise' owing to the numerical discretisation error in the ValoMC simulations.  Bearing these factors in mind, an error saturation at around 5\% in the $\mu_a$ results appears to be reasonable.  We also investigate the influence of the additive Gaussian noise on the errors through including the results obtained without additive noise in Table \ref{tab:Ex1}. For $N\geqslant 3$,  where the error has saturated,  we find that the additive Gaussian noise accounts for between 1.5\% and 3\% of the (approximately) 5\% error figures achieved for the $\mu_a$ reconstructions  with additive noise.  In addition,  the errors for the $\mu_s$ reconstructions do not appear to be strongly influenced by the presence of additive Gaussian noise. This is also consistent with our earlier hypothesis that regularisation is not necessary for this initial study, since we observe that these low noise levels cause only a slight deterioration in the accuracy of our reconstructions, and so it is not necessary to dampen the effect of this noise with regularisation to obtain accurate results for this example.}

\begin{figure}[h!]
\centering
\includegraphics[trim={4cm 0.5cm 3.5cm 0.0cm},clip,width=\textwidth]{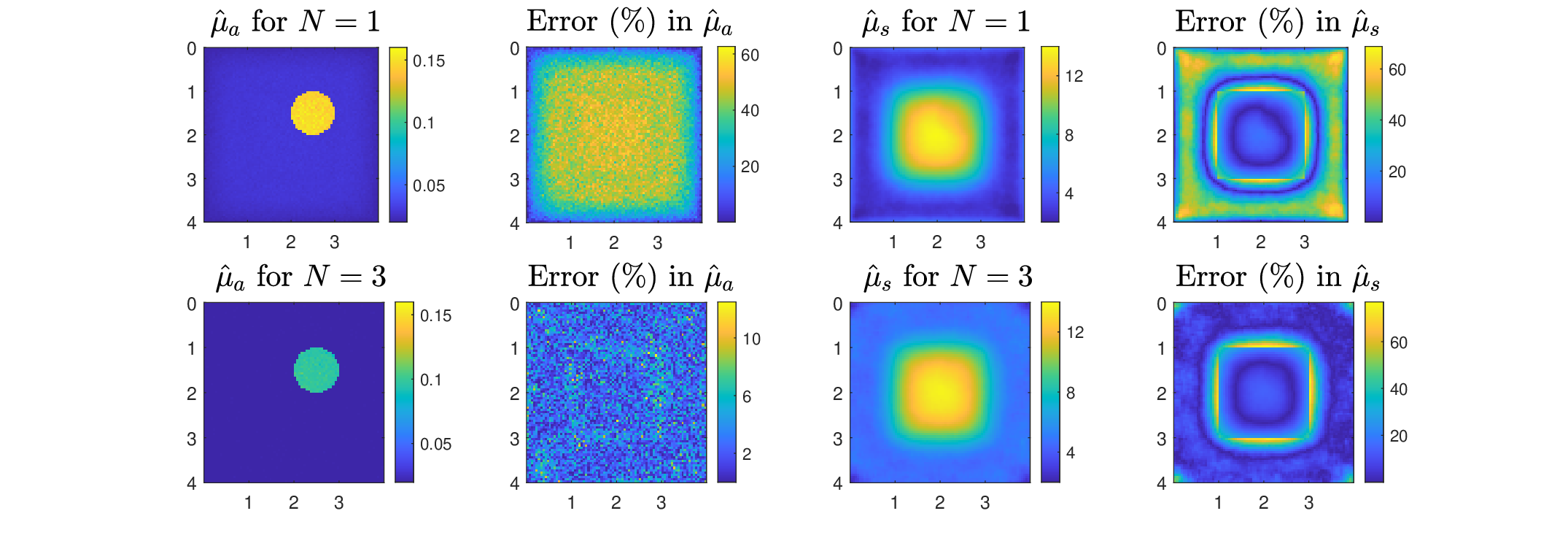}
\caption{\label{fig:ScaledEx1} Values of the absorption (\textcolor{black}{column 1}) and scattering (\textcolor{black}{column 3}) coefficients reconstructed \textcolor{black}{from 400 iterations of the quasi-Newton algorithm} using logarithmically scaled optical energy density data for the initial study, with different truncation values $N$ of the Fourier expansion giving the directional approximation.  \textcolor{black}{Columns 2 and 4 show the (percentage) reconstruction errors in each pixel corresponding to the adjacent result in columns 1 and 3, respectively.} }
\end{figure}
Table \ref{tab:Ex1} also lists the percentage errors $E(\mu_a)$ and $E(\mu_s)$ for the case of logarithmically scaled optical energy density data\textcolor{black}{, both with and without including the 5\% additive Gaussian noise}.  In this case we are able to obtain more accurate reconstructions of $\mu_s$ across all values of $N$ and slightly more accurate reconstructions of $\mu_a$ for $N>2$.  In particular, we note that our $N=3$ result has improved to $E(\mu_a)=3.7\%$ and $E(\mu_s)=16.9\%$ for a relatively modest increase in computational time of around $12.5\%$.  \textcolor{black}{As before,} we do not observe any significant or consistent decrease in either error value for $N>3$ \textcolor{black}{and the additive Gaussian noise accounts for a similar proportion of the total error}.  Figure \ref{fig:ScaledEx1} shows a comparison between the absorption and scattering coefficients reconstructed with $N=1$ and $N=3$ Fourier terms \textcolor{black}{after 400 iterations of the quasi-Newton algorithm using logarithmically scaled noisy data}.  We notice a significant improvement in the previous accuracy issues for the reconstruction of $\mu_s$ close to the region where $\mu_a$ undergoes an abrupt change, which is mirrored by the smaller values of $E(\mu_s)$ shown in Table \ref{tab:Ex1}.  \textcolor{black}{The errors in the reconstruction of $\mu_s$ remain structured however,  with peaks close to the region where $\mu_s$ undergoes an abrupt change in value.  The error distribution in the reconstruction of $\mu_a$ is also similar to the unscaled result,  increasing towards the centre of the domain for $N=1$, but unstructured otherwise. } 

Overall,  our initial study has provided excellent results for the reconstruction of $\mu_a$, and a choice of $N=3$ Fourier terms is recommended to balance computational expense and accuracy.  The relative error values \textcolor{black}{were} around 3-5\% for the synthetic problem data including Gaussian noise scaled to 5\% of the magnitude of the acoustic pressure data\textcolor{black}{,  and only 1-3\% without including additive noise.} The results for the reconstruction of $\mu_s$ were not as accurate with errors in the range $17-20\%$ for $N=3$.  However,  as noted earlier, the chromophore concentrations are linearly related to $\mu_a$ and therefore the quality of the image reconstruction in QPAT crucially depends on the accuracy of the solution for $\mu_a$ only. In both cases, smaller error values were achieved after logarithmically scaling the problem data. The increase in the accuracy provided by logarithmically scaling the data is evident from comparing figures \ref{fig:UnscaledEx1} and \ref{fig:ScaledEx1}, and is obtained with only a modest increase in the associated computational costs.  In the next section, we will extend our analysis to reconstruct a larger and more complex phantom-like image with greater resolution.  

\subsection{Application to reconstruct a phantom-like image}
 \begin{figure}[h!]
\centering
\includegraphics[trim={1.5cm 0.5cm 1.5cm 0.5cm},clip,width=0.75\textwidth]{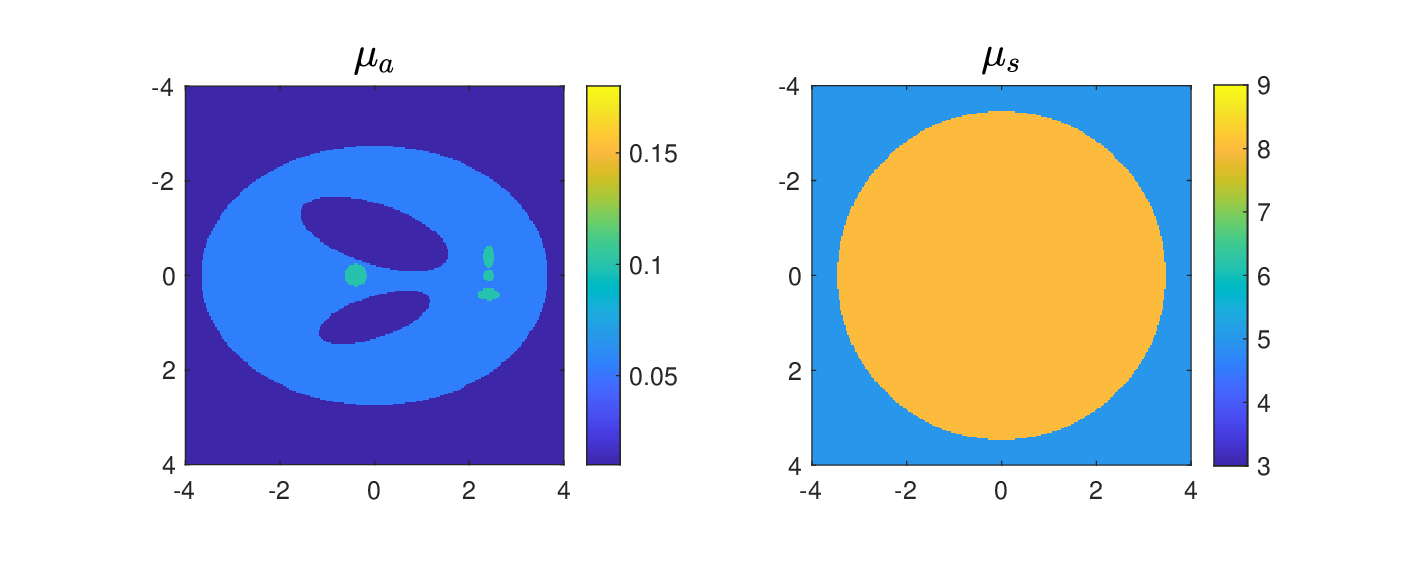}
\caption{\label{fig:GroundTruthEx2} Ground truth values of the absorption (left) and scattering (right) coefficients for the phantom-like image.  The colour axis has been extended for better comparison with the numerically reconstructed absorption and scattering coefficients.}
\end{figure}
We now seek to reconstruct the absorption and scattering coefficients in a 8mm$\times$8mm domain where the background values are taken as $\mu_a=0.01\mathrm{mm}^{-1}$, $\mu_s=5\mathrm{mm}^{-1}$ with anisotropy factor $g=0.8$.  \textcolor{black}{These background values are also used to provide constant initial guesses for $\mu_a$ and $\mu_s$ for the iterative quasi-Newton minimisation scheme.} The ground truth values for $\mu_a$ and $\mu_s$ used in the generation of the synthetic data for a phantom-like image are shown in Figure \ref{fig:GroundTruthEx2}. The images are shown on a finer grid than before with 256$\times$256 pixels, which also corresponds to the number of the pixels $M=65536$ used for the finite volume discretisation of the inverse problem.  We will again investigate the influence of changing the number of Fourier terms $N$, and logarithmically scaling the optical energy density data, on the accuracy of the reconstructions.
 \begin{figure}[h!]
\centering
\includegraphics[trim={4cm 0.5cm 3.5cm 0cm},clip,width=\textwidth]{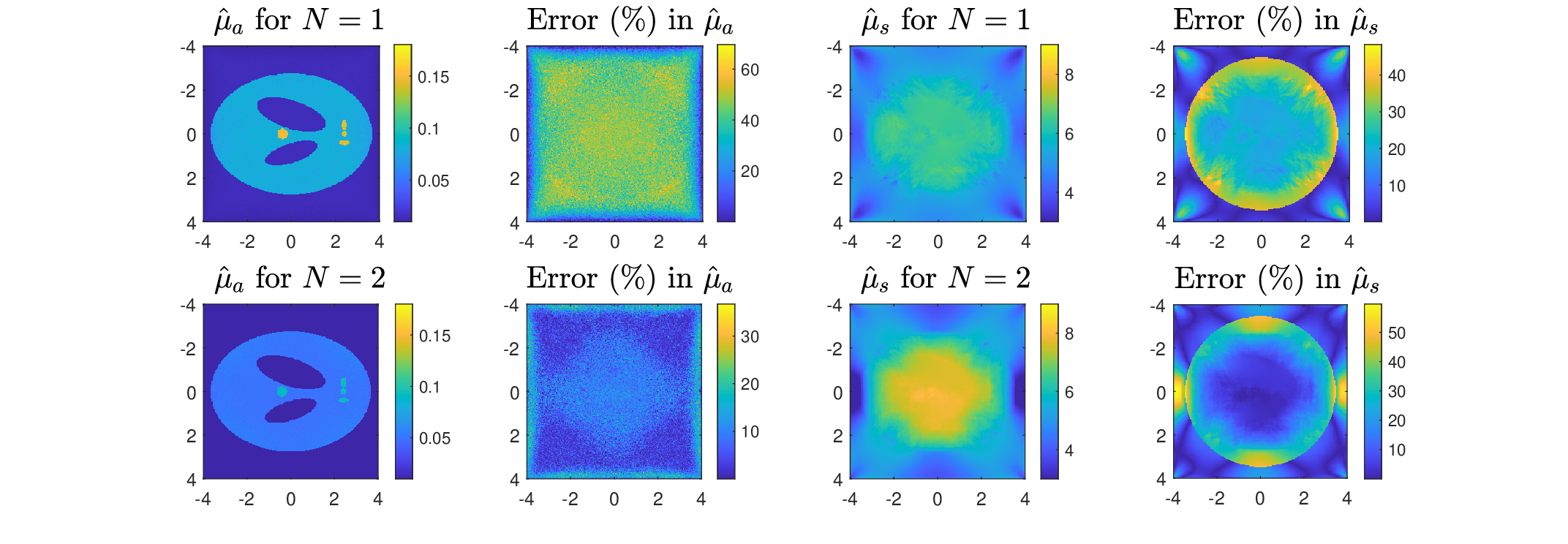}
\caption{\label{fig:UnscaledEx2} Values of the absorption (\textcolor{black}{column 1}) and scattering (\textcolor{black}{column 3}) coefficients reconstructed \textcolor{black}{from 400 iterations of the quasi-Newton algorithm} using unscaled optical energy density data for reconstructing the phantom-like image,  with different truncation values $N$ of the Fourier expansion giving the directional approximation.  \textcolor{black}{Columns 2 and 4 show the (percentage) reconstruction errors in each pixel corresponding to the adjacent result in columns 1 and 3, respectively.}}
\end{figure}

\begin{table}
\caption{\label{tab:Ex2} Percentage errors (\ref{eq:error}) in the \textcolor{black}{non regularised} reconstructions of the absorption and scattering coefficients shown in Figure \ref{fig:GroundTruthEx2} \textcolor{black}{after 400 iterations of the quasi-Newton algorithm} for different truncation values $N$ of the Fourier expansion giving the directional approximation,  and for both unscaled and logarithmically scaled data. \textcolor{black}{Results are also provided for both data with added Gaussian noise at a level of $5\%$ of the acoustic pressure, as well as data with no added noise.}}
\begin{center}
\begin{tabular}{@{}ccccccccccc}
 & \multicolumn{5}{c}{Unscaled data} & \multicolumn{5}{c}{ Logarithmically scaled data}\\
 & \multicolumn{2}{c}{Noisy} & \multicolumn{2}{c}{Clean} & & \multicolumn{2}{c}{Noisy} &\multicolumn{2}{c}{Clean} & \\  
 \cmidrule(r){2-6}  \cmidrule(r){7-11}
$N$ & $E(\mu_a)$ & $E(\mu_s)$ & $E(\mu_a)$ & $E(\mu_s)$ & Time (s) & $E(\mu_a)$ & $E(\mu_s)$ & $E(\mu_a)$ & $E(\mu_s)$ & Time (s)\\ \hline
1 & 44.9 &  24.7 & 44.8 & 25.6 & 6743   & 108 & 26.9 & 57.3 & 23.1   & 858 \\
2 & 8.82 &  20.8 & 7.74 & 21.2  & 17204  & 2350 & 20.0 & 1450 & 21.1   & 25014 \\
3 & 11.0 &  19.4 & 10.4 & 19.5 & 32542  & 2480 & 19.0 & 1830  & 19.3   & 45426 \\
4 & 9.20 &  19.2 & 8.68 & 19.6 & 52414  & 2580 & 18.6 & 2450  & 16.1   & 72289 \\
5 & 8.91 &  19.5 & 8.29 & 19.7 & 84972  & - & - & -   & - & -\\
6 & 9.80 &  19.2 & 8.67 & 18.4 & 114706 & - & - & -   & - & -
\end{tabular}
\end{center}
\end{table}
 
Figure \ref{fig:UnscaledEx2} shows a comparison between the absorption and scattering coefficients reconstructed with $N=1$ and $N=2$ Fourier terms \textcolor{black}{after 400 iterations of the quasi-Newton algorithm using unscaled noisy data, and without applying regularisation (at this stage)}.  One can observe a significant improvement in the reconstructions relative to the ground truth shown in Fig.~\ref{fig:GroundTruthEx2} for both the absorption and scattering coefficients when $N=2$. As with the simple test case studied previously, the reconstruction of $\mu_a$ appears to be superior to the reconstruction of $\mu_s$ \textcolor{black}{and the errors in the latter case are more structured, peaking close to regions where $\mu_s$ undergoes an abrupt change in value.  The error distribution in the reconstruction of $\mu_a$ is again more uniform, having its minimum close to the boundary when $N=1$ and peaking near the boundary when $N=2$.} These observations are quantified in Table \ref{tab:Ex2}, which shows a reduction in $E(\mu_a)$ from \textcolor{black}{$45\%$ to $8.8\%$}, and a reduction in $E(\mu_s)$ from \textcolor{black}{$25\%$ to $21\%$}, when increasing $N$ from 1 to 2.  We note that the accuracy of the reconstruction of the \textcolor{black}{scattering} coefficient is similar to that achieved for the previous simple test case,  however the errors in the reconstructed \textcolor{black}{absorption} coefficient have \textcolor{black}{increased from approximately $5\%$ to around $9\%$.  For $N>2$,  the relative errors in the reconstruction of $\mu_s$ saturate, as before,  at around 19\%. However, the relative errors for $\mu_a$ fluctuate slightly more in the range 9-11\%.  We see a similar pattern in the unscaled data results without additive noise, but with slightly lower errors for the reconstruction of $\mu_a$. We note that the error reduction here is more modest than before, with the error values for the noisy data typically only around $0.5\%$ higher than those achieved without additive noise.  This suggests that the other sources of error (fixed spatial FVM discretisation and discretisation error from ValoMC) are having a greater influence on the results here.} 

The effect of logarithmically scaling the optical energy density data is also shown in the right half of Table \ref{tab:Ex2}.  \textcolor{black}{In this case,  rescaling the data provides a small improvement in the reconstruction accuracy for $\mu_s $, but the reconstruction accuracy for $\mu_a$ has been destroyed.  The results without the Gaussian additive noise are similar, suggesting that the other sources of error are again the dominant influence. Here, rather than improve the accuracy of our reconstructions for $\mu_a$, the logarithmic rescaling has increased the sensitivity of the quasi-Newton algorithm to noise from the ValoMC discretisation error.  In order to improve these results, we will now investigate the effect of including first-order Tikhonov regularisation in the error functional via (\ref{TikReg}).}

\begin{table}
\caption{\label{tab:Ex2r} \textcolor{black}{Percentage errors (\ref{eq:error}) in the regularised reconstructions of the absorption and scattering coefficients shown in Figure \ref{fig:GroundTruthEx2} for different truncation values $N$ of the Fourier expansion giving the directional approximation,  and for both unscaled and logarithmically scaled data with added Gaussian noise at a level of $5\%$ of the acoustic pressure.  The regularisation parameters were taken to be $\alpha=0$, $\beta=7.5\textsc{e}{-10}$ for the unscaled data, and $\alpha=2.5\textsc{e}{-4}$, $\beta=2.5\textsc{e}{-2}$ for the logarithmically scaled data.}}
\begin{center}
\begin{tabular}{@{}ccccccccc}
 & \multicolumn{4}{c}{Unscaled data} & \multicolumn{4}{c}{ Logarithmically scaled data}\\ \cmidrule(r){2-5}  \cmidrule(r){6-9}
$N$ & $E(\mu_a)$ & $E(\mu_s)$ & Iterations &Time (s) & $E(\mu_a)$ & $E(\mu_s)$& Iterations & Time (s)\\ \hline
1 & 48.6 & 33.2 & 20 & 649 & 26.2 & 33.2 & 8 & 858 \\
2 & 5.80 &  32.8 & 20 & 1793 & 9.75 & 33.2 & 15 & 1635 \\
3 & 7.74 & 32.7 & 19 & 3104 & 13.0 & 33.2 & 16 & 3806\\
4 & 7.19 & 32.9 & 19 & 4721 & 13.0  & 33.2 & 18 & 6922\\
5 & 7.25 & 32.7 & 20 & 9605 & 14.6 & 33.2 & 13 & 9281\\
6 & 7.62 & 32.6 & 20 & 10311 & 11.9 & 33.2 & 12 & 10582
\end{tabular}
\end{center}
\end{table}
\textcolor{black}{Table \ref{tab:Ex2r} shows the relative error results obtained after applying first-order Tikhonov regularisation.  We only consider the case of data with additive Gaussian noise here because the noise-free data was included previously to better understand the effect of the additive noise in the absence of regularisation.  Since the smallest error in Table \ref{tab:Ex2} for reconstructing $\mu_a$ was achieved for $N=2$ and this choice  achieves a good balance between accuracy and computational expense,  then we optimise our regularisation parameters through trial and error for the case $N=2$. For the unscaled data, we take $\alpha=0$, $\beta=7.5\textsc{e}{-10}$, and for the logarithmically scaled data we take $\alpha=2.5\textsc{e}{-4}$, $\beta=2.5\textsc{e}{-2}$. With the exception of the $N=1$ result for unscaled data, this choice of regularisation parameters improves the accuracy of the $\mu_a$ reconstruction in all other cases.  In particular, we note the improvement in the error for $N=2$ with unscaled data from $8.8\%$ to $5.8\%$, which is now getting closer to the accuracy achieved for the simpler initial test case.  The accuracy for the logarithmically scaled data is now lower than for the unscaled data due to the more severe ill-posedness of the problem in this case, and for $N=2$ we achieve an error of $9.8\%$.  However, we note that this represents a considerable improvement on the corresponding result from Table \ref{tab:Ex2}. Unfortunately, the $\mu_s$ reconstruction accuracy deteriorates compared to the results without regularisation in Table \ref{tab:Ex2}.  However,  a significant advantage of applying Tikhonov regularisation is that the quasi-Newton minimisation algorithm reaches its stopping criteria before the maximum number of iterations, which in this work has been taken as the 400 iterations quoted for all previous results. The algorithm is stopped by the function tolerance stopping criteria,  which along with all other tolerances (step and optimality) has been set to $10^{-12}$,  significantly smaller than the default value of $10^{-6}$.  The regularised algorithm is now converging after a maximum of 20 iterations (see Table \ref{tab:Ex2r}), which provides a significant order of magnitude saving in the computational run time. }

\begin{figure}[h!]
\centering
\includegraphics[trim={4cm 0.5cm 3.5cm 0cm},clip,width=\textwidth]{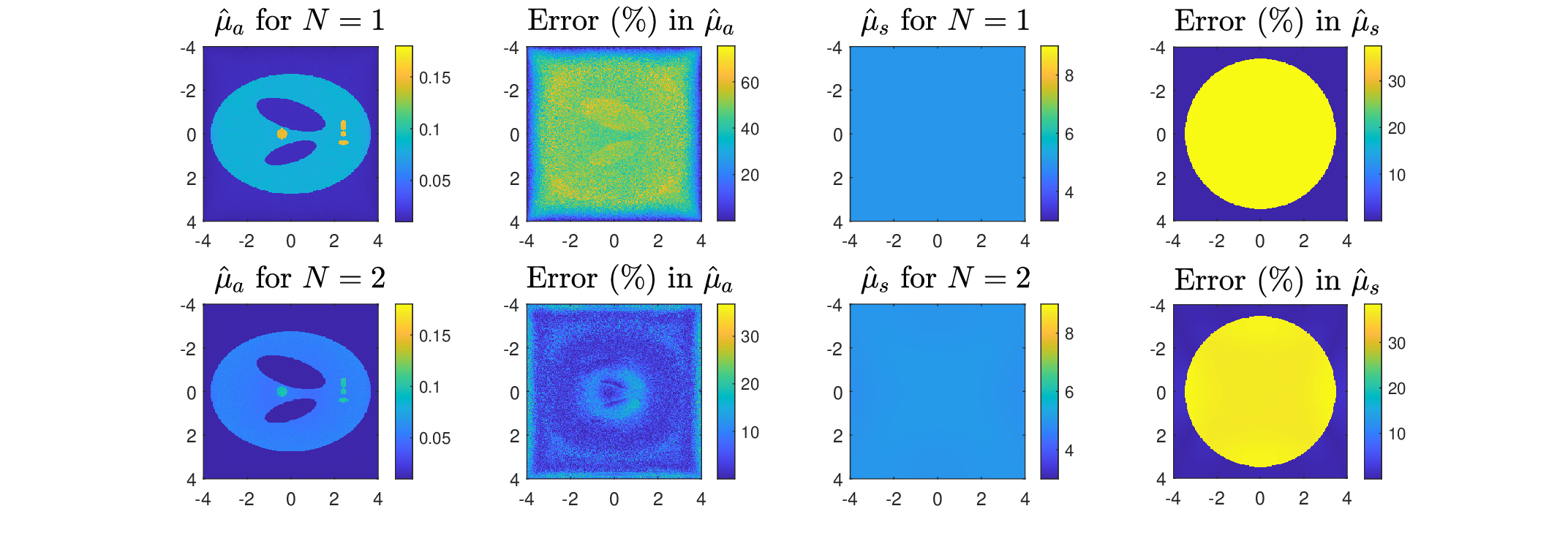}
\caption{\label{fig:RegEx2} Values of the absorption (\textcolor{black}{column 1}) and scattering (\textcolor{black}{column 3}) coefficients reconstructed \textcolor{black}{from 20 iterations of the quasi-Newton algorithm} using \textcolor{black}{unscaled} optical energy density data \textcolor{black}{and first-order Tikhonov regularisation} for reconstructing the phantom-like image,  with different truncation values $N$ of the Fourier expansion giving the directional approximation.  \textcolor{black}{Columns 2 and 4 show the (percentage) reconstruction errors in each pixel corresponding to the adjacent result in columns 1 and 3, respectively.}}
\end{figure}

The \textcolor{black}{regularised} reconstructions for \textcolor{black}{the unscaled data with} $N=1$ and $N=2$ are compared in Figure \ref{fig:RegEx2}. \textcolor{black}{The results show a considerable improvement in the accuracy of the $\mu_a$ reconstruction for $N=2$ compared to $N=1$ and show that the reconstruction for $\mu_s$ has not changed significantly from the initial background value supplied to the quasi-Newton algorithm. The error plots also support these observations, showing that the error in the $\mu_s$ reconstruction is entirely within the circular region where the value differs from the background.  For $\mu_a$,  the error distributions appear to have slightly more structure than before, with some of the geometric features of the phantom visible within the error plot, particularly for $N=1$. Comparing the regularised result for $\mu_a$ with $N=2$ here to the non-regularised equivalent result in Figure \ref{fig:UnscaledEx2}, then it is difficult to see a visible improvement here where the error is lower (5.8\% compared to 8.8\%).}

Overall,  this second example has provided \textcolor{black}{good} results for the reconstruction of $\mu_a$, and a choice of $N=2$ Fourier terms is recommended to balance computational expense and accuracy.  The achieved $E(\mu_a)$ values \textcolor{black}{after applying first-order Tikhonov regularisation} are in the range \textcolor{black}{6-8\% for $N>1$}, which is \textcolor{black}{reasonable} considering the $5\%$ additive Gaussian noise included within the synthetic problem data.  The results for the reconstruction of $\mu_s$ were \textcolor{black}{similar to the initial simple test case for the non-regularised reconstructions,  but did not change significantly from the supplied initial background value in the regularised case.} Since the image reconstruction in QPAT crucially depends on the accuracy of the solution for $\mu_a$, then the gain in the accuracy through reconstructing this quantity using \textcolor{black}{first-order Tikhonov regularisation} is worthwhile\textcolor{black}{, especially considering} the \textcolor{black}{corresponding order of magnitude decrease} in the \textcolor{black}{required} computational time.  

\section{Conclusions}\label{sec:conc}

This paper presents a \textcolor{black}{combined spatial finite volume and directional Fourier-Galerkin} approach for discretising the steady-state RTE formulation of the optical inverse problem of QPAT.  The main advantages of the approach are that the spatial finite volume scheme provides a natural and simple approach for the discretisation of piecewise constant image data and the truncated Fourier expansion in the direction variable means that the method interpolates between the well-known diffusion approximation when $N=1$ and the full RTE model as $N\rightarrow\infty$.  This latter property means that we can easily tune the precision of the model to the demands of the imaging application, taking $N=1$ for cases when the diffusion approximation would suffice and increasing $N$ otherwise.  We made use of the nonlinear optimisation functionality of Matlab to perform gradient based quasi-Newton minimisation via the LBFGS algorithm.  We have included numerical experiments for two test-cases of increasing complexity and resolution, and achieved accurate reconstructions of the absorption coefficient $\mu_a$ for both cases, in the sense that the reconstruction error was \textcolor{black}{of the same order as the} additive noise level in the synthetic problem data.  \textcolor{black}{Depending on the size and complexity of the example, the accuracy of the reconstructed absorption coefficient could be enhanced by either logarithmically rescaling the data or employing first-order Tiknonov regularisation.  In the latter case, which we found appropriate for the larger and more complex example studied,  we also achieved a considerable reduction in the computational run time owing to the earlier convergence of the quasi-Newton minimisation algorithm.} Our chosen examples both demonstrated cases where the diffusion approximation ($N=1$) proved insufficient, and significant accuracy gains were achieved through a modest increase in $N$ up to $N=2$ or $N=3$.  In future work, we would like to expand this promising and flexible modelling approach to larger three-dimensional optical inverse problems using the spherical harmonics in place of the Fourier expansions used here.

\section*{Data availability statement} 
The data that support the findings of this study are available upon reasonable request from the authors.

\section*{Acknowledgments} I would like to thank Professors Ben Cox and Tanja Tarvainen for helpful discussions on the optical inverse problem of QPAT.

\end{document}